\documentclass{SciPost}

\hypersetup{
    colorlinks,
    linkcolor={red!50!black},
    citecolor={blue!50!black},
    urlcolor={blue!80!black}
}

\usepackage[bitstream-charter]{mathdesign}
\DeclareSymbolFont{usualmathcal}{OMS}{cmsy}{m}{n}
\DeclareSymbolFontAlphabet{\mathcal}{usualmathcal}

\fancypagestyle{SPstyle}{
\fancyhf{}
\lhead{\colorbox{scipostblue}{\bf \color{white} ~SciPost Physics }}
\rhead{{\bf \color{scipostdeepblue} ~Submission }}

\fancyfoot[C]{\textbf{\thepage}}
}

\begin{document}

\pagestyle{SPstyle}

\begin{center}{\Large \textbf{\color{scipostdeepblue}{
Exact zero modes in interacting Majorana X- and Y-junctions
}}}\end{center}

\begin{center}\textbf{
Bowy M. La Rivi\`ere\textsuperscript{1$\star$},
Rik Mulder\textsuperscript{1} and
Natalia Chepiga\textsuperscript{2}
}\end{center}

\begin{center}
{\bf 1} Kavli Institute of Nanoscience, Delft University of Technology, Lorentzweg 1, 2628CJ Delft, The Netherlands
\\[\baselineskip]
{\bf 2} Rudolf Peierls Centre for Theoretical Physics, University of Oxford, Clarendon Laboratory, Oxford OX1 3PU, United Kingdom
\\[\baselineskip]
$\star$ \href{mailto:b.m.lariviere@tudelft.nl}{\small b.m.lariviere@tudelft.nl}
\end{center}

\section*{\color{scipostdeepblue}{Abstract}}
\textbf{\boldmath{%
We report the emergence of exact zero modes in junctions of two, three and four short interacting Majorana wires, equivalent to a chain with an impurity bond, Y- and X- junctions respectively. 
These exact zero modes arise from incommensurate short-range correlations induced by interacting Majorana fermions and manifest as exact level crossings between in-gap states upon continuously tuning the interaction strength.
In a junction of only two chains we report exact zero modes and parity switching as soon as the coupling between the chains across a junction is positive.
Remarkably, for junctions with multiple chains the in-gap states group up into sets of parity pairs -- pairs of states with opposite parity and similar energies.
We demonstrate that the formation of these parity pairs are always due to the effective interaction of the outer edges of the junction. 
The behavior within each pair can be efficiently described by two coupled chains. 
In the Y-junction, we detect four in-gap states (two parity pairs) that show exact zero modes not only within each pair but also between them. 
This is attributed to an additional Majorana fermion localized at the center of junction that is protected by symmetry. 
Therefore, coupling between the Majorana fermions at the outer edges of the junction is mediated by that in the center.
We argue that this is a generic feature of junctions with an odd number of arms. 
In the X-junction we detect eight in-gap states (four parity pairs) that are the result of two Majorana degrees of freedom localized at the center of the junction. 
However, we demonstrate that, by contrast to the Y-junction, the appearance of Majorana fermions at the center of the X-junction is not protected and the interaction across the junction can be tuned to the point where there are only Majorana fermions localized at the four outer edges of the junction, forming four in-gap states.
}}

\vspace{\baselineskip}

\noindent\textcolor{white!90!black}{%
\fbox{\parbox{0.975\linewidth}{%
\textcolor{white!40!black}{\begin{tabular}{lr}%
  \begin{minipage}{0.6\textwidth}%
    {\small Copyright attribution to authors. \newline
    This work is a submission to SciPost Physics. \newline
    License information to appear upon publication. \newline
    Publication information to appear upon publication.}
  \end{minipage} & \begin{minipage}{0.4\textwidth}
    {\small Received Date \newline Accepted Date \newline Published Date}%
  \end{minipage}
\end{tabular}}
}}
}


\vspace{10pt}
\noindent\rule{\textwidth}{1pt}
\tableofcontents
\noindent\rule{\textwidth}{1pt}
\vspace{10pt}


\section{Introduction}\label{sec: Introduction}
Topological phases of matter have been at the forefront of condensed matter physics, with a particular interest in Majorana fermions -- particles that are their own antiparticles.
As noted by Kitaev \cite{kitaevUnpairedMajoranaFermions2001a}, 
a chain of spinless Dirac fermions, an effective model of a $p$-wave superconductor, displays topoligically non-trivial properties with a single "unpaired" zero-energy Majorana fermion localized at its edges.
This generates a partner-state with opposite parity for each state in the many-body spectrum, creating a degeneracy for each of them.
If this affects the entire spectrum these zero energy modes are referred to as \textit{strong} zero modes (SZM) -- a term popularized by Fendley \cite{fendleyParafermionicEdgeZero2012a}.
Because of their foundational framework for fault tolerant quantum computing \cite{RevModPhys.80.1083,sarmaMajoranaZeroModes2015,marraMajoranaNanowiresTopological2022}, 
observing signatures of these SZM has been the hallmark in various experimental works over the past decades \cite{aliceaNewDirectionsPursuit2012,beenakkerSearchMajoranaFermions2013,dasZerobiasPeaksSplitting2012,dengAnomalousZeroBiasConductance2012,mourikSignaturesMajoranaFermions2012,rokhinsonFractionalAcJosephson2012,toskovicAtomicSpinchainRealization2016}.

In the Kitaev chain Majorana fermions are non-interacting, but there are multiple proposals extending the model to the interacting case \cite{fidkowskiEffectsInteractionsTopological2010,fidkowskiTopologicalPhasesFermions2011,rahmaniEmergentSupersymmetryStrongly2015,rahmaniPhaseDiagramInteracting2015,katsuraExactGroundStates2015,verresenStableLuttingerLiquids2019c,chepigaTopologicalQuantumCritical2023a}.
Such interacting Majorana chains have been studied in various contexts, including quantum phase transitions \cite{selaMajoranaFermionsStrongly2011,chepigaCriticalPropertiesMajorana2023a,chepigaEightvertexCriticalityInteracting2023,chepigaTopologicalQuantumCritical2023a,laflorencie2023universalsignaturesmajoranazero}, and in the presence of quenched disorder \cite{crepinNonperturbativePhaseDiagram2014,gergsTopologicalOrderKitaev2016,karcherDisorderInteractionChiral2019,lobosInterplayDisorderInteraction2012,milstedStatisticalTranslationInvariance2015,robertsInfiniteRandomnessContinuously2021,chepigaResilientInfiniteRandomness2024a} with a recent focus on many-body localization at high energy \cite{kjallManyBodyLocalizationDisordered2014,laflorencieTopologicalOrderRandom2022,moudgalyaPerturbativeInstabilityDelocalization2020,sahayEmergentErgodicityTransition2021,wahlLocalIntegralsMotion2022}.
Furthermore, Majorana edge states have been shown to survive in the presence of interactions, with studies demonstrating both their robustness and the interaction-induced modifications to their properties \cite{stoudenmireInteractionEffectsTopological2011,hasslerStronglyInteractingMajorana2012,thomaleTunnelingSpectraSimulation2013,zvyaginChargingMajoranaEdge2022}.
Several forms of interaction have been proposed, but here we focus on the Hamiltonian with the simplest non-trivial translationally-invariant interaction term, described by
\begin{equation}
\label{eq: interacting kitaev}
    \mathcal{H} = 
    i \sum_{j=1}^{N-1} t_j \gamma_{j} \gamma_{j+1}
    - \sum_{j=1}^{N-3} g_j \gamma_{j} \gamma_{j+1} \gamma_{j+2} \gamma_{j+3},
\end{equation}
where $\gamma_j$ are the typical Majorana fermionic operators satisfying $\gamma_j^\dagger = \gamma_j$, $\gamma_j^2=1$ and $\{ \gamma_i, \gamma_j \} = 2 \delta_{ij}$; $N$ the number of Majorana fermions in the chain; the staggered hopping amplitudes and coupling strengths are denoted by $t_{2j}$, $t_{2j+1}$ and $g_{2j}$, $g_{2j+1}$ respectively, corresponding to the even and odd components.

Contrary to the non-interacting case, no explicit and exact construction of SZM exist for this model -- or interacting chains in general \cite{kellsManybodyMajoranaOperators2015,kellsMultiparticleContentMajorana2015,wieckowskiIdentificationMajoranaModes2018,komaStabilityMajoranaEdge2022}, with the notable exception being the integrable XXZ chain \cite{fendleyStrongZeroModes2016,PhysRevLett.133.050606}.
However, in incommensurate phases the effective coupling between edges states can be fined-tuned to a set of points where it vanishes and, correspondingly, the energy levels of the in-gap states cross each other -- called {\it exact} zero modes (EZM) \cite{toskovicAtomicSpinchainRealization2016,chepigaExactZeroModes2017a,vionnetLevelCrossingsInduced2017,wadaCoexistenceStrongWeak2021,alexandradinataParafermionicPhasesSymmetry2016}.
While these do not guarantee the presence of SZM, EZM do signal edge states.
On top of that, in-gap states are easily accessible numerically through exact diagonalization or by targeting excited states in the density matrix renormalization group algorithm \cite{chepigaExcitationSpectrumDensity2017}.

At the same time, interest in the physics of junctions in general has steadily increased following technological developments in coupled superconducting wire systems \cite{PhysRevB.77.155422,PhysRevB.79.085122,GIULIANO2009395,PhysRevB.85.045120}, as well as in the context of non-Abelian braiding protocols and network architectures based on tri-junctions of Majorana wires \cite{aliceaNonAbelianStatisticsTopological2011,sauControllingNonAbelianStatistics2011,clarkeMajoranaFermionExchange2011,vanheckCoulombassistedBraidingMajorana2012,beenakkerSearchMajoranaFermions2013,aliceaNewDirectionsPursuit2012a}.
Furthermore, recent conceptual progress on Majorana corner states in second-order topological superconductors \cite{pahomiBraidingMajoranaCorner2020,zhangTopologicalHolonomicQuantum2020,ikegayaTunableMajoranaCorner2021,yanMajoranaCornerModes2018} that may eventually open up new avenues to explore more complex junction architectures.
However, in most of these settings the focus lies on idealized non-interacting or effectively non-interacting regimes.
In particular, the outer edge states are either screened \cite{white-Yjunction} or pushed far from the junction \cite{BUCCHERI201552}, or the Majorana fermions are non-interacting \cite{pandeyMajoranaZeroModes2023}.
As a result, a systematic study of localized edge states in multi-chain junctions of interacting Kitaev chains remains comparatively limited.
Here we aim to fill this knowledge gap by studying Majorana edge states through EZM in interacting Kitaev chains coupled at the edges in moderately sized Y- and X-junctions.

\begin{figure}[!b]
  \begin{center}
    \includegraphics{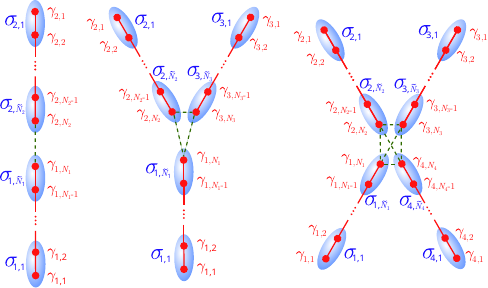}   
    \caption{
        Sketches of the junction geometries addressed in this paper; two coupled Majorana chains (left), Y-junction (middle), and X-junction (right).
        Dark green dotted lines denote the coupling between the ends of the chains that is described by $\mathcal{H}_\mathrm{coupling}(\alpha)$.
        In red we mark the Majorana degrees of freedom $\gamma_{i,k}$ in arm $i$, with $ 1 \leq k \leq N_i$. 
        And in dark blue the corresponding spin-1/2 $\sigma_{i,k}$, with $1 \leq k \leq \tilde{N}_i = N_i/2$. 
    }
    \label{fig: illustrations}
  \end{center}
\end{figure}

In this paper we focus on the family of interacting Majorana chains defined in Eq.\eqref{eq: interacting kitaev} coupled at one edge, as illustrated in Fig.\ref{fig: illustrations}, by some coupling term $\mathcal{H}_\mathrm{coupling}(\alpha)$, with $\alpha$ being an external parameter controlling the strength of the coupling across the junction. 
For practical reasons we re-formulate the Hamiltonian of Eq.\eqref{eq: interacting kitaev} in terms of spins using the Jordan-Wigner transformation (see Appendix \ref{Ap: A} for more details):
\begin{equation}\label{eq: full Hamiltonian spin}
  \mathcal{H}^{(n)} = 
  \sum_i^{n}
    \left(
      -h \sum_{j=1}^{\tilde{N}_i} 
      \sigma_{i,j}^z 
      + \sum_{j=1}^{\tilde{N}_i-1} \left(
        J\sigma_{i,j}^x \sigma^x_{i,j+1} 
      + g \sigma_{i,j}^z \sigma_{i,j+1}^z \right)
      + g \sum_{j=1}^{\tilde{N}_i-2} \sigma_{i,j}^x \sigma_{i,j+2}^x 
    \right)
    + \mathcal{H}_\mathrm{coupling}(\alpha),
\end{equation}
where $\sigma_{i,j}^x$ and $\sigma_{i,j}^z$ are Pauli matrices acting on site $j$ in chain $i$ and $n$ is the number of chains.
Each chain $i$ contains $\tilde{N}_i = N_i/2$ spins.
We use $h$ and $J$ notations of the transverse field Ising model, $t_{2j+1}=h$ and $t_{2j}=-J$, explicitly allowing alternation of the Majorana hopping term when $h \neq J$, and we set $g_{2j}=g_{2j+1}=g$.
$\mathcal{H}_\mathrm{coupling}(\alpha)$ depends on the geometry of the junction and will be defined explicitly in the corresponding sections.
We limit ourselves to the symmetric case in which $J$, $g$, $h$ and $N$ are the same for each chain, and we assume an even number of Majorana fermions per chain.

We study oscillations in the in-gap energy spectrum by tuning the magnetic field $h$ -- effective parameter that allows tuning of the coupling between the edge states in this model -- while initializing the system in the incommensurate topologically nontrivial parity-broken $\mathbb{Z}_2$ phase, located at $|h| < 1$ and interaction strengths $0 < g \lesssim 0.4$ \cite{chepigaTopologicalQuantumCritical2023a}.
The chosen parameter regime does not cross the Ising critical point at the non-interacting point $(h,g)=(1,0)$, nor the corresponding critical line at finite $g$.
We fix $g=0.2$ unless stated otherwise.
For more information on the calculation of the low-energy spectrum of the Hamiltonian, we refer to Appendix \ref{Ap: B}.


Before proceeding, we briefly identify the symmetry class of the constituent chains. The non-interacting Kitaev chain ($g=0$) preserves fermion parity and an effective time-reversal symmetry, placing it in symmetry class BDI and yielding a $\mathbb{Z}$ topological classification with two Majorana end modes \cite{PhysRevB.78.195125,ryuTopologicalInsulatorsSuperconductors2010}. In the presence of a four-Majorana interaction ($g \neq 0$), this effective time-reversal symmetry is broken down to fermion parity alone \cite{fidkowskiEffectsInteractionsTopological2010,fidkowskiTopologicalPhasesFermions2011}, reducing the classification to class D with a $\mathbb{Z}_2$ invariant.
This $\mathbb{Z}_2$ structure of class D persists upon coupling multiple Kitaev chains into a junction and constrains the number of unpaired Majorana modes.
In a junction of $n$ chains, each contributing two boundary Majorana modes, a total of $n$ Majorana degrees of freedom meet at the junction.
For odd $n$, one unpaired Majorana mode is symmetry-protected, as we will illustrate for the Y-junction.
For even $n$, all Majorana modes can, in principle, pair into $n/2$ Dirac fermions without symmetry obstruction, as exemplified by the two coupled chains studied below.
However, our numerics for the X-junction show that microscopic details of the system can lead to additional Majorana modes remaining at the junction beyond this minimal symmetry-based counting.

The rest of the paper is structured as follows.
In Sec. \ref{sec: Two coupled chains} we benchmark our method with the simplest case of two coupled chains. 
We present numerical evidences of the exact zero modes in the Y-junction in Sec. \ref{sec: Y-junction}. 
In Sec. \ref{sec: X-junction} we study exact zero modes appearing in X-junction and discuss the strategy to manipulate them.
Finally, we conclude our results and put them into perspective in Sec. \ref{sec: conclusion}.


\begin{figure}[!b]
  \begin{center}
    \includegraphics{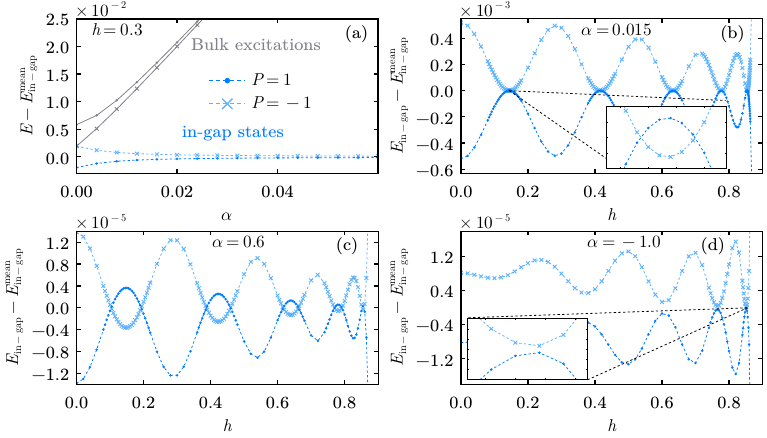}   
    \caption{
      Numerical results for the energy spectrum of two chains of $N=20$ Majorana fermions each, coupled by Eq.\eqref{eq: 2-chain coupling}.
      We fix the four-Majorana coupling strength at $g=0.2$. 
      All data are centered around the average of the two lowest energy levels.
      (a) Four lowest energy levels as a function of the coupling strength $\alpha$. The magnetic field $h=0.3$ is fixed.
      When coupling the two chains ($\alpha > 0$), two states with opposite parity (light blue crosses and dark blue dots) form the in-gap spectrum while the other two become part of the bulk excitations (grey lines).
      (b) In-gap spectrum as a function of the magnetic field $h$ for a small coupling strength $\alpha=0.015$.
      Exact zero modes appear as soon as $\alpha>0$. 
      For small $\alpha$ the intervals where negative parity sector forms the ground-state are much shorter compared to that of positive one.
      Inset: first crossings between the parity sectors zoomed in.
      (c) Same as in (b) but for $\alpha=0.6$. 
      Interval in which the ground state has a negative parity is larger compared to $\alpha=0.015$. 
      (d) Same as in (b) but for negative coupling $\alpha=-1$.
      We observe no parity switching and no exact zero modes.
      Inset: Final avoided crossing zoomed in.
    }
    \label{fig: two chains EZM}
  \end{center}
\end{figure}

\section{Two coupled chains}\label{sec: Two coupled chains}
We start our analysis with the simplest case of two chains coupled to each other.
Equivalently, one can consider this system as a long chain with a single impurity bond.
This system is similar to the junction studied recently in Ref.\cite{PhysRevB.111.L201114}, in which two non-interacting Kitaev chains with different magnetic fields $h$ and coupling strengths $J$ were considered. 
Here we consider the coupling of two identical but interacting Majorana chains instead.
We choose the coupling term such that when the junctions coupling strength $\alpha \to 1$ we retrieve a single interacting Kitaev chain.
Therefore, the coupling term consists of all two and four Majorana operators that cross the junction, coupling the outer three Majorana fermions of both chains according to
\begin{equation}\label{eq: 2-chain coupling}
    \begin{aligned}
        \mathcal{H}_\mathrm{coupling}(\alpha) = & \mathcal{H}^{(i,j)}\\ 
        \equiv & \alpha \left[
            - J i \gamma_{i,N_i} \gamma_{j,N_j} 
            - g (
            \gamma_{i,N_i-2} \gamma_{i,N_i-1} \gamma_{i,N_i} \gamma_{j,N_j} \right. \\
            & \quad + \gamma_{i,N_i-1} \gamma_{i,N_i} \gamma_{j,N_j-1} \gamma_{j,N_j}
            \left. +\gamma_{i,N_i} \gamma_{j,N_j-2} \gamma_{j,N_j-1} \gamma_{j,N_j}  
            )
        \right],
    \end{aligned}
\end{equation}
where $i$ and $j$ label two chains, simply $i=1$ and $j=2$ here.
We parametrize the two-Majorana term by the hopping amplitude $-J$, in accordance with Eq.\eqref{eq: full Hamiltonian spin}, since it couples Majorana fermions located on 'adjacent' spin sites.
Through the Jordan-Wigner transformation, we find that
\begin{equation}\label{eq: 2-chain coupling spin}
    \mathcal{H}^{(i,j)} = 
      \alpha g \sigma_{i,\tilde{N}_i}^z \sigma_{j,\tilde{N}_j}^z
      + \alpha i \left(
        J \sigma^x_{i,\tilde{N}_i} \sigma^x_{j,\tilde{N}_j}  
        + g \sigma_{i,\tilde{N}_i-1}^x \sigma_{j,\tilde{N}_j}^x
        + g \sigma_{i,\tilde{N}_i}^x \sigma_{j,\tilde{N}_j-1}^x
      \right) 
      \prod_{i < m \leq j, k=1 }^{k=\tilde{N}_m} \sigma_{m,k}^z.
\end{equation}
The string of $\sigma^z$ operators is a result of the two chains being ordered such that its ends meet.
Note, one of the $\sigma^z$ operators in this string acts on the same site as the $\sigma^x$ operator in chain $j$ (effectively hiding the Hermitian nature of the coupling at a first glance).
For two chains of $N_i=N_j=20$ Majorana fermions coupled by Eq.\eqref{eq: 2-chain coupling}, we show the four lowest energies as a function of the coupling strength $\alpha$ in Fig.\ref{fig: two chains EZM}(a) centered around the average of the two smallest energies for visual clarity.
We label each energy level through the parity 
\begin{equation}
    P = \prod_{i=1}^{n}\prod_{j=1}^{j=\tilde{N}_i} \sigma_{i,j}^z
\end{equation}
of its corresponding state, which we compute by taking the product over all spins in the system.
For a single finite-sized interacting Kitaev chain, the in-gap spectrum consists of a ground state and a quasi-degenerate excited state with opposite parity.
Two decoupled copies of this chain therefore yield four in-gap states: a unique ground state and a second excited state, both with $P=1$, as well as a twofold-degenerate first excited state with $P=-1$. 
As expected, we observe this spectrum at the decoupled point $\alpha=0$.
When we couple the two chains this degeneracy is lost and two of the four in-gap states with opposite parity pair up -- from now on we will refer to such a pair as {\it parity pairs} -- and form the new in-gap spectrum.
The other two states become part of the bulk excitation spectrum.

\begin{figure}[!b]
  \begin{center}
    \includegraphics[width=0.7\textwidth]{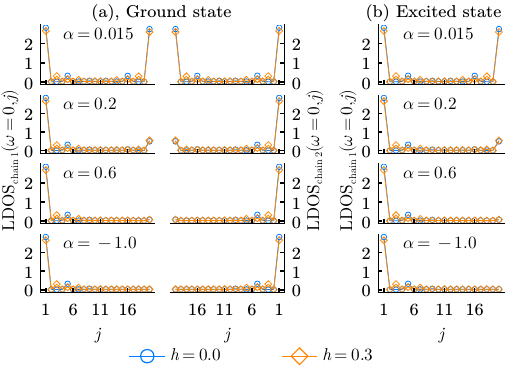}   
    \caption{
      Spatial profile two chains containing $N=20$ Majorana fermions each, coupled by four different strengths $\alpha$.
      Data is provided for the two in-gap states.
      Four-Majorana interaction strength is fixed at $g=0.2$.
      Blue circles and orange diamonds depict the strength of the magnetic field $h$.
      (a) Spatial profile for the ground state.
      Left side indicates the first chain while the right side the second chain.
      Amplitude of local density of states for $\alpha=0.015$ in the middle of the chain is slightly smaller than at the edges.
      As expected, the local density of states in the center vanishes when approaching the single interacting Kitaev chain limit, i.e. $\alpha \rightarrow 1$.
      For negative coupling $\alpha=-1$ the Majoranas in the center is also absent.
      (b) Same as in (a) but for the excited state.
      Local density of states is only shown for the first arm. 
      Note that the profile is visually identical to that of the ground state.
    }
    \label{fig: two chains LDOS}
  \end{center}
\end{figure}

For the in-gap spectrum we show its dependence on the magnetic field $h$ in Fig.\ref{fig: two chains EZM}(b) for two weakly coupled chains, i.e. $\alpha=0.015$.
There are clear oscillations with a vanishing energy gap between the two parity sectors.
Unlike the uniform interacting Kitaev chain though, the interval in which the negative parity sector is the ground state is much smaller than that of a positive parity sector.
As demonstrated in Fig.\ref{fig: two chains EZM}(c), upon increasing the coupling up to $\alpha=0.6$ the intervals over which the negative parity sector is the ground state become much wider.
Intuitively, this is a result of the two coupled chains approaching the single-chain limit $\alpha\rightarrow 1$ where symmetry between even and odd parity sectors is restored.

We also computed the in-gap spectrum for the case of negative coupling $\alpha = -1$ across the junction, presented in Fig.\ref{fig: two chains EZM}(d).
Now we detect a set of avoided crossings without any parity switching, and thus no EZM appear.
This provides a first indication that the sign of the interaction at a junction plays a crucial role on the formation of EZM.

In addition to the energy spectrum, we also visualize the spatial location of Majorana fermions in the two chains by calculating the local density of states (LDOS)   
\begin{equation}\label{eq: LDOS}
    \mathrm{LDOS}^{(l)}_k (\omega,j) = \mathrm{Im}
        \left[
            \left\langle \psi_l \left |
            \gamma_{k,j} \frac{1}{\omega + i \eta - \mathcal{H}^{(n)} + E_l} \gamma_{k,j}
            \right| \psi_l \right\rangle
        \right],
\end{equation} 
for each \textit{Majorana} fermion $1 \leq j \leq N_k$ in chain $k$.
For both in-gap states $\psi_l$, with an energy $E_l$, we carry out this procedure for the two chains in the junction.
We always set the broadening parameter $\eta = 0.1$ and compute the LDOS at frequency $\omega = 0$ -- value for which a Majorana zero mode is normally expected.
We calculate the LDOS numerically by iteratively solving the eigenvalue problem obtained from Eq.\eqref{eq: LDOS} with the conjugate-gradient method (see \cite{kuhnerDynamicalCorrelationFunctions1999} for more detail).
In Fig. \ref{fig: two chains LDOS}(a) we provide the LDOS for the ground state for four different coupling strengths $\alpha$ and two magnetic fields $h$.
In all four cases there are clear Majorana edge states localized at the ends of the two chains.
There also appear two modes at the center of the junction for small $\alpha$ that slowly disappear upon increasing $\alpha$ and upproaching a single-chain limit.
The LDOS profiles for $\alpha=-1$ and $\alpha=+1$ are similar.
Interestingly, as show in Fig.\ref{fig: two chains LDOS}(b), the spatial profile of both in-gap states is nearly identical, highlighting that the states differ solely by the interaction between Majorana edge states rather than by their spatial appearance and location.


\section{The Y-junction: three coupled chains}\label{sec: Y-junction}
Let us now move on to the case when three chains are coupled at the ends, forming the Y-junction.
The full three-chain coupling term
\begin{equation}\label{eq: 3-chain coupling full}
    \begin{aligned}
        \mathcal{H}_\mathrm{coupling}(\alpha) = 
        \left( \sum_{i=1}^{2}\sum_{j > i}^{3} \mathcal{H}^{(i,j)} \right)
        +  \mathcal{H}^{(1,2,3)},
    \end{aligned}
\end{equation}
consists of all possible two chain couplings described by Eq.\eqref{eq: 2-chain coupling} (the terms in the brackets) and the term that couples three chains labeled $i$, $j$ and $k$:
\begin{equation}\label{eq: 3-chain coupling}
        \mathcal{H}^{(i,j,k)} \equiv
        - g \alpha^2 
        \left(
            \gamma_{i,N_i-1} \gamma_{i,N_i} \gamma_{j,N_j} \gamma_{k,N_k}
            + \gamma_{i,N_i} \gamma_{j,N_j-1} \gamma_{j,N_j} \gamma_{k,N_k}
            +\gamma_{i,N_i} \gamma_{j,N_j} \gamma_{k,N_k-1} \gamma_{k,N_k}  
        \right).
\end{equation}
This term follows naturally from the construction outlined in the previous section.
The factor $\alpha^2$ stems from the Majorana operators crossing the gap between the chains twice.
In terms of spin-operators
\begin{equation}\label{eq: 3-chain coupling spin}
    \begin{aligned}
        \mathcal{H}^{(i,j,k)} = & g \alpha^2 i \left( 
            \sigma^z_{i,\tilde{N}_i} \sigma_{j,\tilde{N}_j}^x\sigma_{k,\tilde{N}_k}^x \prod_{j< m \leq k, l = 1}^{l=\tilde{N}_m}\sigma_{m,l}^z \right. \\
            & \quad\quad\quad\quad + \sigma_{i,\tilde{N}_i}^x \sigma^z_{j,\tilde{N}_j} \sigma_{k,\tilde{N}_k}^x \prod_{i < m \leq k, l=1}^{l=\tilde{N}_m}\sigma^z_{m,l} \\
            & \left. 
            \quad\quad\quad\quad  + \sigma_{i,\tilde{N}_i}^x \sigma_{j,\tilde{N}_j}^x \sigma_{k,\tilde{N}_k}^z \prod_{i < m\leq j, l=1}^{l=\tilde{N}_m} \sigma^z_{m,l} \right).
    \end{aligned}
\end{equation}
Again, the $\sigma^z$ strings are due to the chosen ordering of the chains.

For a junction of three chains containing $N=14$ Majorana fermions each, coupled by Eq.\ref{eq: 3-chain coupling spin}, we show the eight lowest energy levels as a function of the coupling strength $\alpha$ in Fig.\ref{fig: Y-junction EZM}(a) (we provide the same plot but for the full range $-1 \leq \alpha \leq 1$ in Appendix \ref{Ap: Y-junction alpha}). 
The data is centered around the average of the in-gap spectrum for finite $\alpha$ for visual clarity. 
At the decoupled point $\alpha=0$ the degenerate sets of excitations come in triplets that is a direct consequence of having three arms in the junction. 
The ground state corresponds to the state with all three arms of the junction being in the parity sector $P=1$; the first three-fold degenerate excitation set has $P=-1$ and corresponds to two arms with parity $1$ and one arm with parity $-1$; the next set has two arms with parity $-1$ and one with parity $+1$, so the overall parity is $P=1$ again; finally, the eighth state, highest in energy, corresponds to all three arms being in the parity $-1$ sector.
As soon as $\alpha>0$ four out of the eight states become part of the bulk excitation spectrum while the remaining four make up the in-gap spectrum and form two parity pairs. 
For comparison, we provide the results for $N=12$ Majorana fermions per arm featuring similar exact zero modes in Appendix \ref{Ap: Y-junction fewer sites}.

\begin{figure}[t]
  \begin{center}
    \includegraphics[width=\textwidth]{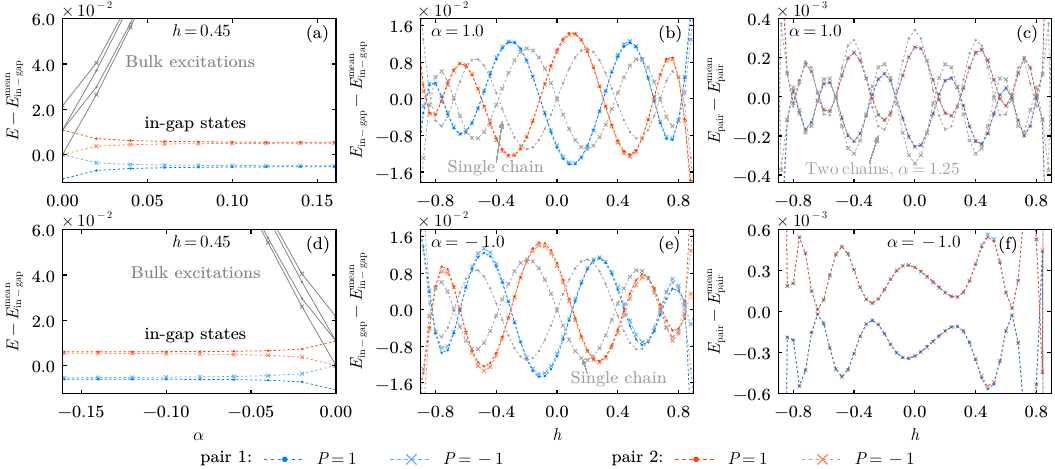}
    \caption{
      Energy spectrum of three interacting Kitaev chains, each containing $N=14$ Majorana fermions, coupled as defined by Eq. \ref{eq: 3-chain coupling full} in a Y-junction geometry for a fixed four-Majorana interaction strength $g=0.2$.
      Dots and crosses denote the positive and negative parity sectors respectively.
      (a) Eight lowest energies, centered around the average of the four in-gap states, as a function of the coupling strength $\alpha$ and a fixed magnetic field $h=0.45$.
      In a Y-junction four states make up the in-gap spectrum and group up into two parity pairs (red and blue curves), while the other four become bulk excitations (grey lines).
      (b) In-gap spectrum as a function of $h$ at $\alpha=1$, centered around the average of the lowest four levels.
      Note that there are two levels marked with red and two different levels with blue that are barely distinguishable on this scale; the oscillations within each pair is analyzed in (c).
      Parity pairs show multiple crossings as a fincvtion of $h$. For reference, we show level crossings in a single interacting Kitaev chain of $N=14$ Majorana fermions (grey curve).
      (c) Same data as in (b) but centered around the average of each parity pair.
      For comparison, we show two chains of $N=14$ Majorana fermions each coupled with a strength $\alpha=1.25$.
      (d)-(f) In-gap spectrum for $\alpha=-1$ plotted with  the same convention as in (a)-(c). While we still see two parity pairs (d) and crossing between them (e), the crossings within each pair are now avoided (f).
    }
    \label{fig: Y-junction EZM}
  \end{center}
\end{figure}

\begin{figure}[t]
  \begin{center}
    \includegraphics[width=0.7\textwidth]{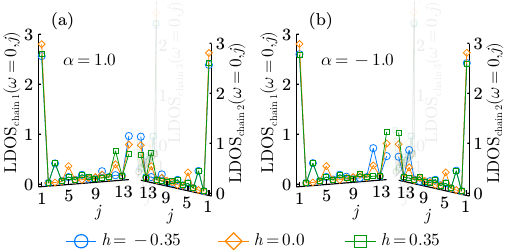}   
    \caption{
      Spatial profile of Majorana fermions in the Y-junction for two coupling strenghts $\alpha$.
      Data is presented for the ground state -- for other in-gap states we refer to Appendix \ref{Ap: Y-junction LDOS in-gap states} -- and for three different magnetic fields $h$ (blue circles, orange diamonds and green squares).
      The four-Majorana interaction strength is fixed at $g=0.2$.
      (a) Local density of states for all three chains for $\alpha=1$.
      Third chain is greyed out for visual clarity.
      Each chain has a clear Majorana zero mode localized at its edge and another mode appears at the center of the junction with its amplitude distributed equally among the three arms.
      The amplitude and spatial distribution varies with the magnetic field.
      Asymmetry in the energy spectrum with respect to the magnetic field (see Fig.\ref{fig: Y-junction EZM}(b)) is also present here.
      (b) Same as in (a) but for $\alpha=-1$.
      Note, the spectrum appears the same as in (a) but mirrored around $h=0$. 
    }
    \label{fig: Y-junction LDOS}
  \end{center}
\end{figure}

We show the dependence of this in-gap spectrum on the magnetic field $h$ for a fixed $\alpha=1$ in Fig.\ref{fig: Y-junction EZM}(b).
Again, for visual clarity, we center the data around the average of the four energy levels. 
Similar to the previous case we detect exact level crossings with a vanishing energy gap.
Note, this time it is not the energy within each parity pairs that is vanishing but the energy gap between the two pairs.
Since the splitting between the parity sectors vanishes fast with the system size\footnote{
For instance, there are two consecutive crossing points located within the distance $\delta h\approx 2\cdot 10^{-3}$ for $N=10$ Majorana fermions per arm.
Increasing the number of Majorana sites to $N=14$ per arm reduces this the distance to $\delta h\approx 2\cdot 10^{-4}$.
}, 
all four states are degenerate at the points of exact zero modes.
As a reference, we also show the two in-gap states of a single interacting Kitaev chain of the same length as a single arm in the Y-junction (grey lines in Fig.\ref{fig: Y-junction EZM}(b)).
Quite fascinating, the single chain shows the same number of crossings between the two parity sectors as the ones occurring between two parity pairs of the Y-junction. 
This implies that the crossings we observe are controlled by a characteristic length of a single arm, rather then, for instance the interaction between the outer edges separated by a lattice distance that is twice as large. 
This interesting phenomenon is attributed to the odd number of Majorana zero modes at the center of the junction.
Since they can not annihilate all at once, an additional Majorana degree of freedom is localized at the center of the junction and the crossing {between the pairs} are due to the interaction of this additional Majorana zero mode with those localized at the outer edges.
Therefore, we expect similar behavior for other multi-chain junctions with an odd number of arms.

In Fig.\ref{fig: Y-junction EZM}(c) we plot the same data as in the panel (b) but now centered around the average of each parity pair.
There is a stark symmetry present: the same parity sectors of the two parity pairs completely mirror each other\footnote{Red crosses almost perfectly overlap with blue dots, and visa versa}.
Strikingly, the oscillations resemble those of two chains of $N=14$ Majorana fermions coupled by a strength $\alpha=1.25$. 
In Appendix \ref{Ap: Y-junction vary g} we show that this similarity with this non-universal value of $\alpha$ holds throughout most of the incommensurate region. 
Intuitively, one can argue that in the Y-junction each arm is coupled to two other arms and that it is therefore natural to expect $\alpha>1$ when comparing with two coupled chains.
This simple comparison demonstrates that characteristic length scale for level crossings within each parity sector is twice larger than the length of an arm implying that the crossings within each parity pair are, by contrast to the previous case, due to interactions of the Majorana fermions localized at the outer edges. 
This picture is further supported by twice as many crossings in Fig.\ref{fig: Y-junction EZM}(c) compared to Fig.\ref{fig: Y-junction EZM}(b), and a significantly smaller amplitude that is expected to decay exponentially fast with the distance between the localized edge states.

When the chains are coupled with a negative coupling $\alpha<0$ we again report four in-gap states that group into two parity pairs, as shown in Fig.\ref{fig: Y-junction EZM}(d).
Like in the case with positive coupling $\alpha$ we see crossings between the two parity pairs when we vary the magnetic field $h$ (see Fig.\ref{fig: Y-junction EZM}(e)) with a characteristic length scale similar to that of a single arm.
In striking contrast is the behavior within each parity pair presented in Fig.\ref{fig: Y-junction EZM}(f) where we detect no parity switching and thus no exact level crossings.

We would like to emphasize that the difference between the cases of $\alpha$ being $-1$ or $1$ is interesting.
When centering the data around the entire in-gap spectrum, both cases show similar behavior. 
The similarity with a single chain that is the length of a single arm suggests that the dominant energy scale here is that of the coupling between the outer edges and the Majorana in the center of the junction.
By centering the data around each parity pair we effectively 'zoom-in' on the coupling between the outer edges. 
This is evident from the reduced amplitude and increased frequency of the oscillations, characteristic of systems where the interaction between the edge states can be continuously tuned \cite{chepigaExactZeroModes2017a,toskovicAtomicSpinchainRealization2016}. 
However, only when the coupling is positive do we observe a vanishing of the parity gap.

We provide the spatial profile of the Majorana fermions in the junction for the ground state in Fig.\ref{fig: Y-junction LDOS}(a) for $\alpha = 1$ and three values of the magnetic field $h$ (we show the LDOS for the full in-gap spectrum in Appendix \ref{Ap: Y-junction LDOS in-gap states}).
Each of the arms is symmetric with respect to one another and has a clear Majorana zero mode localized at the outer edges -- similar to two couple chains.
On top of this, another Majorana degree of freedom is localized at the center of the junction that is equally distributed between the three chains.
One can notice that for $h>0$ the Majorana degree of freedom in the center of the junction is less localized, effectively shortening the distance between the Majorana fermions at the edges and the center.
In Fig.\ref{fig: Y-junction LDOS}(b) we show the LDOS $\alpha=-1$.
The profile is qualitatively similar to that of $\alpha>0$: a single Majorana degree of freedom localized in the center of the junction and three localized at the outer edges.


\section{The X-junction: four coupled chains}\label{sec: X-junction}

Let us now proceed with the X-junction in which we couple the ends of four chains.
The Hamiltonian describing this coupling is given by 
\begin{equation}\label{eq: X-junction}
    \mathcal{H}_\mathrm{coupling}(\alpha) = 
    \left( \sum_{i=1}^{3}\sum_{j > i}^{4} \mathcal{H}^{(i,j)} \right)
    + \left( \sum_{i=1}^{2}\sum_{j > i}^{3} \sum_{k > j}^{4} \mathcal{H}^{(i,j,k)} \right)
    + \mathcal{H}^{(1,2,3,4)},
\end{equation}
where the terms in the first brackets contain all possible two chain couplings, the second all possible three chain couplings, described by Eq.\eqref{eq: 2-chain coupling} and Eq.\eqref{eq: 3-chain coupling} respectively, and the final term 
\begin{equation}\label{eq X-junction coupling}
    \begin{aligned}
        \mathcal{H}^{(i,j,k,l)} \equiv & -g \alpha^3 \gamma_{i,N_i} \gamma_{j,N_j} \gamma_{k,N_k} \gamma_{l,N_l} \\
        = & -g \alpha^3 \sigma^x_{i,\tilde{N}_i} \sigma^x_{j,\tilde{N}_j} \sigma^x_{k,\tilde{N}_k} \sigma^x_{l,\tilde{N}_l} 
        \left( \prod_{i<m \leq j, q = 1}^{l = \tilde{N}_q} \sigma_{m,q}^z \right)
        \left( \prod_{k < m \leq l, q = 1}^{q = \tilde{N}_l} \sigma_{m,q}^z \right) 
    \end{aligned}
\end{equation}
couples the outer four Majorana fermions in four chains labeled $i$, $j$, $k$ and $l$.
The factor $\alpha^3$ comes from the Majorana operators crossing the center of the junction three times.
We provide the sixteen lowest eigenvalues for four chains of $N=10$ Majorana fermions in Fig.\ref{fig: X-junction EZM}(a)\footnote{Results for smaller system sizes are qualitatively similar and provided in Appendix \ref{Ap: X-junction fewer sites}}.
When the chains are uncoupled, i.e. $\alpha=0$, both the ground and sixteenth state are unique and have parity $P=1$, in which the former has all arms in the party $+1$ sector while the latter has parity $-1$ for all four of them.
The remaining states group up into three degenerate sets; two of these are quartets with a parity of $-1$ that correspond to either the four possible configurations with a single excitation in an arm, or the four possibilities for an excitation in three arms at once; the other set is six-fold degenerate and accounts for two excitations distributed over two of the four arms.
As soon as the chains are coupled the degeneracy of these states is lifted and eight states form the in-gap spectrum and partner up into four parity pairs while the other eight become part of the bulk excitation spectrum.

\begin{figure}[!t]
  \begin{center}
    \includegraphics[width=0.8\textwidth]{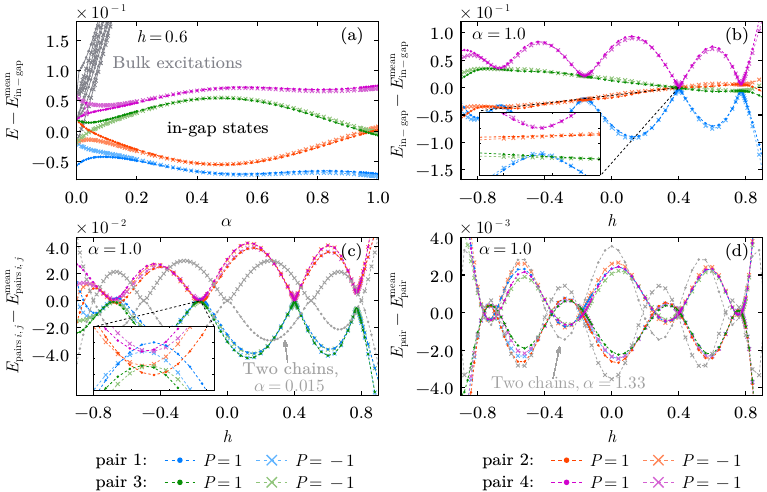}   
    \caption{
        Energy spectrum of four Majorana chains of $N=10$ Majorana fermions each coupled in the X-junction geometry. The  interaction strength is kept constant at $g=0.2$.
        Dots and crosses mark parity sectors $P=1$ and $P=-1$ respectively.
        (a) 16 lowest energy levels as a function of the coupling strength $\alpha>0$, eight of which form four parity pairs (blue, red, green and purple) of in-gap states.
        Magnetic field is kept constant at $h=0.6$.
        Data is centered around the average of these eight in-gap states.
        (b) In-gap spectrum centered around its average as a function of magnetic field $h$.
        Note that there are two levels marked for each of the four pairs that are barely distinguishable on this scale; the oscillations within each pair are analyzed in (d).
        Spectrum looks remarkably different from the Y-junction (see Fig.\ref{fig: Y-junction EZM} for comparison).         
        First and last parity, and third and fourth pair never cross. 
        There is only a single crossing between the second and third pair and two very rapid intersections between the third and fourth pair occur on the right.
        Crossings between first and second pair occur shortly after each other.
        Inset: avoided crossing zoomed between first and fourth pair zoomed in.
        (c) Same data as in panel (b) the group of lower four and higher four in-gap states centered around their corresponding average.        
        Only crossings between the first and second pair occur.
        Inlet: second crossing zoomed in.
        Spectrum acts surprisingly similar to a system of two weakly coupled chains, i.e. by a strength $\alpha=0.015$, of $N=5$ Majoranas each (grey lines).
        (d) Same data as in (b) but centered around the mean of each parity pair.
        Oscillations in the energy show a remarkable similarity with two chains coupled with a coupling strength $\alpha=1.33$ (grey line).
    }
    \label{fig: X-junction EZM}
  \end{center}
\end{figure}

Fig.\ref{fig: X-junction EZM}(b)-(d) shows how the relative energy of these eight in-gap states changes with the magnetic field $h$ for a fixed junction coupling $\alpha=1$.
The picture is more complex and remarkably different in comparison to the Y-junction shown in Fig.\ref{fig: Y-junction EZM}. 
To understand the underlying physics, let us first take a look at the crossings within each parity pair as displayed in Fig.\ref{fig: X-junction EZM}(d). 
As in the case of Y-junction these crossings are reasonably well described by two coupled chains -- this similarity holds through a large part of the incommensurate interval as shown in Appendix \ref{Ap: X-junction LDOS appendix}. 
This implies that the energy splitting within each parity pair, as in all previous cases, are due to the coupling between Majorana fermions localized at the outer edges of the system that are separated by the lattice distance twice larger than the distance of a single arm.

By looking at Fig.\ref{fig: X-junction EZM}(c) we see that crossings between pairs one and two resembles a weakly coupled pair of chains (grey lines)\footnote{The two weakly coupled chains do not fit the first and second parity pairs of the X-junction due to the slight delocalization of the Majorana fermions at the center of the junction, shortening the effective length scale over which the center and edges couple.}, except now we have two copies of it. 
The number of pairwise crossings and periodicities of oscillations in Fig.\ref{fig: X-junction EZM}(c) matches those in Fig.\ref{fig: X-junction EZM}(d). 
At the same time, the other two parity pairs show avoided crossings, as is clearly visible in the inset of Fig.\ref{fig: X-junction EZM}(c), and resembles the physics of two chains with $\alpha<0$ presented in Fig.\ref{fig: two chains EZM}(d), except again we have two copies of each state.

\begin{figure}
  \begin{center}
    \includegraphics[width=\textwidth]{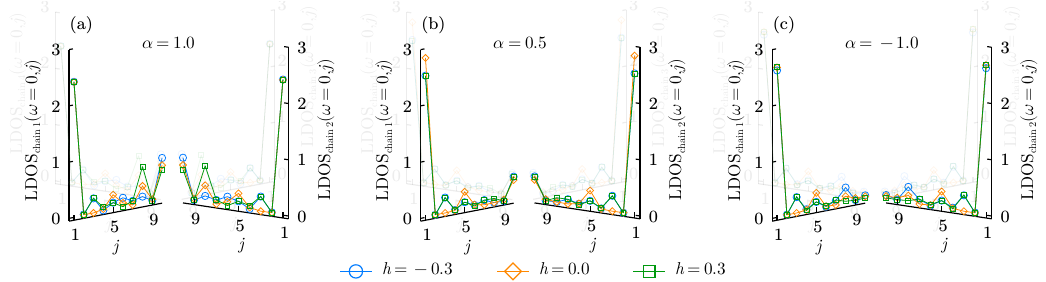}   
    \caption{
        Spatial profile of Majorana degrees of freedom in the X-junction for three different junction strengths; (a) $\alpha=1$, (b) $\alpha=0.5$ and (c) $\alpha=-1$; and three values of the magnetic fields $h$ (blue circles, orange diamonds and green squares); and $g=0.2$. 
        All data are computed for $N=10$ Majoranas per arm and is presented for the ground state only -- data for all in-gap states is provided in Appendix \ref{Ap: X-junction LDOS appendix}. 
        Third and fourth chain are greyed out for visual clarity. 
        In all profiles we see Majorana zero modes localized at the outer edges ($j=1$). 
        In (a) and (b) two Majorana modes are localized at the center of the junction equally distributed between four legs. 
        In (c) We see no signature of the Majorana degrees of freedom at the center of the junction.
        }
    \label{fig: X-junction LDOS}
  \end{center}
\end{figure}

We interpret these results as follows. 
In the center of the X-junction four Majorana fermions neither fully annihilate as in the case of two chains, nor create some symmetric structure with all four degrees of freedom present. 
Instead, our results suggest that a pair of Majorana fermions is localized at the center of the junction. 
This picture is fully supported by our LDOS calculations presented in Fig.\ref{fig: X-junction LDOS}(a) showing that the total density of the Majorana degrees of freedom in the center is about twice as large as the LDOS at each of the four outer edges.
This picture also explains the unusual number of in-gap states that we observe: six Majorana fermions have a total Hilbert space of size $(\sqrt{2})^{6}=8$ and follow from four localized at the outer edges and two in the center of the junction.

\begin{figure}[!t]
    \includegraphics[width=\textwidth]{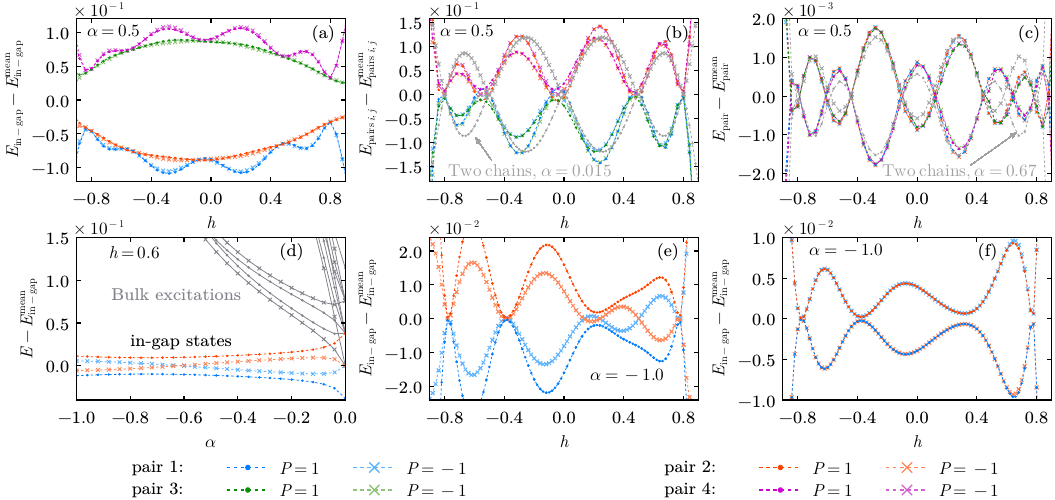}   
    \caption{
        Energy spectrum of four chains of $N=10$ Majorana fermions each coupled in the X-junction geometry. All shown examples are for the interaction strength $g=0.2$.
        Colors denote the different parity pairs that make up the in-gap spectrum; dots (crosses) mark a parity $P=1$ ($P=-1$).
        (a) Eight states that make up the in-gap spectrum for the junction coupling $\alpha=0.5$ as a function of magnetic field $h$. Data is centered around its average.
        Each parity pair contain two energy levels that are visually indistinguishable on this scale. 
        We show the oscillations within each pair in (c).
        (b) Same data as in panel (a) but pair one and two centered around their average, and the same for pair three and four.
        Grey lines depict the in-gap spectrum of a two weakly coupled chains of $N=5$ Majoranas each, which shows a remarkable similarity to the behavior of the first and second parity pairs.
        Note that the two parity sectors of each pair, i.e. the dots and crosses, are visually indistinguishable.
        (c) Same data as in (a) but centered around the mean of each parity pair.
        Oscillations in the energy resemble those of the re-scaled spectrum of two chains coupled with a coupling strength $\alpha=0.67$ (grey line). 
        (d) Spectrum of the sixteen lowest energies as a function of the coupling strength $\alpha \leq 0$ and $h=0.6$.
        By contrast to $\alpha>0$ (see Fig.\ref{fig: X-junction EZM}(a) for comparison) there are only four in-gap states forming two parity pairs. Data is centered around the lowest four states.
        (e) Four in-gap states as a function of the magnetic field $h$ for $\alpha=-1$.
        Positive parity sectors never cross while the negative ones show multiple points with a vanishing energy gap.
        (f) Energy for each parity pair centered around its average. Similar to the the Y-junction for $\alpha=-1$ (Fig.\ref{fig: Y-junction EZM}(f)) we see no parity switching within the pairs.
    }
    \label{fig: X-junction EZM 2}
\end{figure}

In the picture outlined above the state of the pair of Majorana fermions in the center of the junction determines the splitting between the lower and higher four in-gap states. 
At $\alpha=1$ the system is fine-tuned (see Fig.\ref{fig: X-junction EZM}(a)) such that we observe all of them on a roughly same scale. 
For smaller values of $\alpha$, we see a clear separation between the two parity-pair below and the two parity-pairs above, as clearly visible in Fig.\ref{fig: X-junction EZM 2}(a) for $\alpha=0.5$.
When we vary $h$, as shown in Fig.\ref{fig: X-junction EZM 2}(a)-(c), the picture still holds though: despite the four lower states being well separated from the upper four, pairwise crossing (avoided crossings for the upper four levels), as well as crossing within each parity pair, remains qualitatively identical to the case described in Fig.\ref{fig: X-junction EZM}. 
In the LDOS presented in Fig.\ref{fig: X-junction LDOS}(b) we also notice that the Majorana fermions are now better localized in the center of the junction.

The situation is drastically different if we change the parity of the coupling in the junction, examples of which we show in Fig.\ref{fig: X-junction EZM 2}(d)-(f) for $\alpha<0$. 
In Fig.\ref{fig: X-junction EZM 2}(d) we observe only four in-gap states, indicating that the Majorana degrees of freedom in the center of the junction vanished, leaving only the four Majorana edge states localized at the outer edges of the junction. 
This is fully supported by the LDOS calculations that we present in Fig.\ref{fig: X-junction LDOS}(c), with a negligibly small density of states in the center of the junction.
Unlike the Y-junction, the in-gap spectrum behaves qualitatively different from $\alpha>0$ under tuning of the magnetic field $h$, see Fig.\ref{fig: X-junction EZM 2}(e).
Negative coupling at the junction leads to a sequence of level crossings for $h>0$ but only between states with negative parity, and a set of avoided crossings for $h<0$. 
This rich behavior suggests that in junctions with an even number of arms the appearance and character of exact zero modes can be manipulated by the type and strength of the coupling across the junction.
In agreement with our results for $\alpha=-1$ in the Y-junction, presented in Fig.\ref{fig: Y-junction EZM}(f), there are only avoided crossings within each parity pair, suggesting that this feature is generic for all junctions with $\alpha<0$.

\section{Conclusion and Discussion}\label{sec: conclusion}

To summarize, we report the appearance of exact zero modes in interacting Kitaev chains coupled in Y- and X-junctions as well as in a junction coupling two chains through a weaker link.
We focussed on topologically non-trivial phases with alternating hopping between even and odd pairs of Majorana sites and the simplest non-trivial interaction $g$ spanning four consecutive Majorana fermions.
We studied these junctions numerically with exact diagonalization by first mapping them to their dual spin-1/2 models.

In the junction with two chains we report exact zero modes and parity switching in the in-gap spectrum as soon as the coupling across the junction $\alpha$ becomes positive. 
Exact zero modes must also be expected in more generic situations including, in particular, quench disorder with eventual weak bonds as impurities.
This is because the short-range incommensurability that gives rise to the sequence of exact zero modes, by continuously tuning an external parameter, persists in the presence of this disorder \cite{chepigaResilientInfiniteRandomness2024a,2025arXiv250611985S}.
This creates an inspiring perspective to observe regular parity switching in disordered interacting Majorana chains, which we leave for future investigations.

In the Y-junction we report four in-gap states that group into two parity pairs. 
These four states are attributed to there being four Majorana degrees of freedom: three localized at the outer edges and a single Majorana fermion localized in the center of the junction\footnote{In principle, we do not exclude the possibility of having three Majorana degrees of freedom at the center as well for different couplings between the chains}.
This central degrees of freedom mediates the interaction between the outer degrees of freedom and is responsible for the exact level crossings between two parity pairs.
Appearance of the Majorana fermion in the center of the Y-junction is protected by symmetry, and we expect, for the same reason, similar behavior for junctions with an odd number of arms. 
Such junctions can then, as this suggests, be effectively used as a single tuning knob to manipulate all outer edge states. 
This might be of particular interest as the Y-junction has been used to realize braiding of Majorana fermions that requires the constant tuning of variables, such as the electrostatic potential, to move Majorana fermions around in the junction \cite{aliceaNonAbelianStatisticsTopological2011,flensbergNonAbelianOperationsMajorana2011,hellCouplingBraidingMajorana2017,hyartFluxcontrolledQuantumComputation2013,karzigOptimalControlMajorana2015,pluggeMajoranaBoxQubits2017,vanheckCoulombassistedBraidingMajorana2012}.

The X-junction, on the other hand, has eight in-gap states that group into four parity pairs, which stems from the emergence of two Majorana fermions in the center of the junction. 
These degrees of freedom are stabilized by the symmetric four-Majorana coupling term shown in Eq.\eqref{eq X-junction coupling}.
Unlike the Y-junction, the appearance of these two additional Majorana fermions in the center of the junction is not protected:  we have shown that localization of these degrees of freedom and even their appearance can be tuned by the coupling at the junction. 
On top of that, the asymmetric nature of the in-gap spectrum between sets of parity pairs suggest that the X-junction for positive coupling can be viewed as two weakly coupled chains crossing each other in the middle, with the control of the impurity bond depending inversely on the strength of the coupling in the junction.

Finally, we observe that in all junctions in-gap states appear as parity pairs. 
Within each parity pairs the energy levels behave similarly to the two weakly coupled chains, showing exact zero modes for $\alpha>0$ and a set of avoided crossings for $\alpha<0$. 
The characteristic length controlling the frequency of the parity switching is well described by the lattice distance between two outer edges of the junction.
This suggests that the formation of parity pairs are due to interactions between the Majorana fermions localized at those edges. 
This property seems to be generic for any junction and shows no signature of even-odd effect with respect to the number of chains coupled in the junction; neither is it sensitive to the location of the Majorana degrees of freedom near the center at the junction.

\section*{Acknowledgements}
NC thanks to Nicolas Laflorencie and Frederic Mila for insightful discussions and inspiring work on related subjects.

\paragraph{Funding information}
This research has been supported by Delft Technology Fellowship. Numerical simulations have been performed with the Dutch national e-infrastructure with the support of the SURF Cooperative and at the DelftBlue HPC.

\newpage
\begin{appendix}
\numberwithin{equation}{section}

\begin{figure}[!b]
  \begin{center}
    \includegraphics{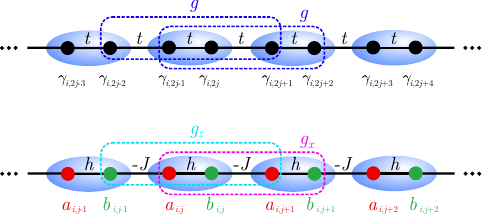}
    \caption{
        Schematic of labeling of the Majorana chain. 
        In the top image we show the ordinary interacting Kitaev chain consisting of Majorana operators $\gamma_{i,2j+1}$ with the hopping $t$ and four-Majorana interaction strength $g$.
        Bottom: same as in the top but all odd $2j+1$ and even $2j$ numbered Majoranas are re-labeled as $a_j$ and $b_j$ respectively.
        In addition, we renamed $t$ and $g$ into the coupling strength $J$, magnetic field $h$ and $x$ and $z$ component of $g$ according to the following: $t(2j) = h$, $t(2j+1) = -J$, $g(2j) = g_z$ and $g(2j+1) = g_x$.
    }
    \label{fig:hamiltonian_explanation}
  \end{center}
\end{figure}

\section{Mapping Majorana fermions to spins}\label{Ap: A}
To map Majorana fermions to spins, we first start of with the usual relation between Dirac fermion and Majoranas:
\begin{equation}\label{eq: Majorana fermions to Dirac fermions}
    \begin{aligned}
        \gamma_{2j-1} &= c_j^\dagger + c_j\\
        \gamma_{2j} &= i(c_j^\dagger - c_j).
    \end{aligned}
\end{equation}
The Dirac fermions can directly be mapped to Pauli spin operator through the Jordan-Wigner transformation, such that
\begin{equation}\label{eq: Jordan-Wigner transformation}
    \begin{aligned}
        \sigma_j^z &= 1 - 2c_j^{\dagger}c_j,\\
        \sigma_j^x &= K_j ( c_j^{\dagger} + c_j ),\\
        \sigma_j^y &= i K_j ( c_j^{\dagger} - c_j ),
    \end{aligned}
\end{equation}
where $K_j = \prod_{k=1}^{j-1}\sigma_k^z$.
For convenience, we rename every odd numbered Majorana operators $\gamma_{2j-1} = a_j$ and all even numbered $\gamma_{2j}=b_j$ (see Fig.\ref{fig:hamiltonian_explanation} for a sketch).
In terms of the odd and even Majorana operators, the Jordan-Wigner transformation can be formulated as 
\begin{equation}
    \begin{aligned}
        a_j &= K_j \sigma_j^x,\\
        b_j &= K_j \sigma_j^y\\
            &= -i K_{j+1} \sigma_{j}^x,\\
        a_j b_j &= i \sigma_j^z,
    \end{aligned}
\end{equation}
where in the second relation we used $\sigma^y = -i\sigma^z \sigma^x$.
Before mapping the interacting Kitaev chain to its dual spin model, we first rewrite Eq. \ref{eq: interacting kitaev} in terms of $a_j$ and $b_j$ as well, such that  
\begin{equation}
    \begin{aligned}
        \mathcal{H} =& i h \sum_{j=1}^{\tilde{N}} a_j b_j - \sum_{j=1}^{\tilde{N}-1} \left( iJ b_j a_{j+1} + g_z a_j b_j a_{j+1} b_{j+1} \right)
        -  g_x \sum_{j=1}^{\tilde{N}-2} b_j a_{j+1} b_{j+1} a_{j+2}.
    \end{aligned}
\end{equation}
Note, the sum runs over the number of Majorana pairs $\tilde{N} = N/2$ -- equivalently the number of spins.
The hopping amplitude $t$ and coupling strength $g$ are renamed into their respective spin counterpart for which $t_{2j+1} = h$, $t_{2j} = -J$, $g_{2j+1} = g_z$ and $g_{2j+1} = g_x$.
We also show this renaming in Fig.\ref{fig:hamiltonian_explanation}.
By applying the Jordan-Wigner transformation to the interacting Kitaev chain we find the spin Hamiltonian
\begin{equation}
    \mathcal{H} =   - h \sum_{j=1}^{\tilde{N}} \sigma_{j}^z 
    + \sum_{j=1}^{\tilde{N}-1} \left( J \sigma_{j}^x \sigma^x_{j+1} + g_z \sigma_{j}^z \sigma_{j+1}^z\right)
    + \sum_{j=1}^{\tilde{N}-2} g_x \sigma_{j}^x \sigma_{j+2}^x,
\end{equation}
which is the as the one shown in the brackets in Eq.\eqref{eq: full Hamiltonian spin} when setting $g_x=g_z=g$.

We always arrange the chains inward, as shown in the illustrations in Fig.\ref{fig: illustrations}, causing the final three Majorana fermions of each chain to be $b_{i,\tilde{N}_i-1}$, $a_{i,\tilde{N}_i}$ and $b_{i,\tilde{N}_i}$.
As result, even for the benchmark case of coupling two chains, strings of $\sigma_z$-operators appear after applying the Jordan-Wigner transformation on the terms $\mathcal{H}_\mathrm{coupling}(\alpha)$ that couple these chains.

\newpage

\section{Computing the low-energy spectrum}\label{Ap: B}
To compute the eigenvalues and eigenvectors of the spin Hamiltonians we use the Lanczos algorithm \cite{lanczosIterationMethodSolution1950} with its subspace spanned by the Krylov basis.
We construct this Krylov basis without explicitly constructing the Hamiltonian as a (sparse) matrix.
For an $N$-body system, any arbitrary many-body quantum state is encoded as a $2^N$ dimensional vector, where each element describes a product state of the $N$ spins and its relative phase in the full many-body state.
Essentially, all the $2^N$ indices of this vector can be straightforwardly converted to a binary string of length $N$.
By considering this as a product state of spin -- we adopt the convention that 1 corresponds to spin-up and 0 to spin-down -- operators can be directly applied to this bit string.
To be more precise, the operators $\sigma^x_j$ and $\sigma^z_j$ act on such a binary string in the following way: $\sigma^x_j$ flips 0 to 1 at site $j$, or visa versa, and $\sigma^z_j$ multiplies the relative phase by $-1$ when in site $j$ is spin down.
Applying the Hamiltonian would entail summing over all product states, or equivalently binary numbers, and tracking how the operators that make up the Hamiltonian affect it.

To illustrate this, we encode $\frac{3}{5} \left|\uparrow \downarrow \right\rangle + \frac{4}{5}\left|\uparrow \uparrow \right\rangle$ as $\left( 0, 0, \frac{3}{5}, \frac{4}{5} \right)^T$, and applying the operator $\sigma^x_0 \sigma^z_1$ to this vector results in $\left( -\frac{3}{5}, \frac{4}{5}, 0, 0 \right)^T$. 

As an initial guess for the Lanczos algorithm we always use a vector with its elements initialized as random complex numbers.

\newpage

\section{Additional results Y-junction}\label{Ap: C}
\subsection{In-gap spectrum as a function of $ -1 \leq \alpha \leq 1$}\label{Ap: Y-junction alpha}

\begin{figure}[!h]
  \begin{center}
    \includegraphics{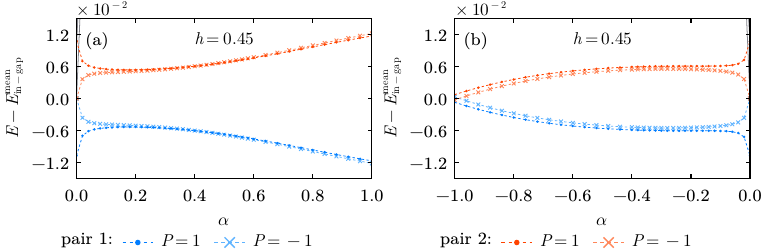}   
    \caption{
        In-gap spectrum of the Y-junction for three chains of $N=14$ Majoranas s a function of the coupling strength $\alpha$.
        Four out of the eight states form the in-gap spectrum and group into two pairs with opposite parity (blue and red lines).
        Crosses depict a parity $P=-1$ and dots $+1$.
        We fix the four-Majorana interaction strength $g=0.2$.
        The spectra are provided for a fixed magnetic field $h=0.45$. 
        (a) Spectrum for $0 \leq \alpha \leq 1$.
        (b) Same as in (a) but for $-1 \leq \alpha \leq 0$.
    }
    \label{fig: Y-junction EZM appendix vary alpha}
  \end{center}
\end{figure}

In the main text we discussed the energy spectrum of the Y-juncton as a function of the coupling strength $\alpha$, see Fig.\ref{fig: Y-junction EZM}(a) and (d).
In Fig.\ref{fig: Y-junction EZM appendix vary alpha} we show the same data but over an extended range $\alpha$, i.e. $0 \leq \alpha \leq 1$ in panel (a) and $-1 \leq \alpha \leq 0$ in panel (b).
The data is shown for a fixed magnetic field $h$; different values of $h$ could result in different curves, but the number of in-gap states remains invariant.  
We observe that the number of in-gap states remains four in both intervals, further confirming the robustness of the four Majorana modes in the junction.

\newpage

\subsection{In-gap spectrum as a function of $h$ for $N=12$ Majoranas per arm}\label{Ap: Y-junction fewer sites}

\begin{figure}[!h]
  \begin{center}
    \includegraphics{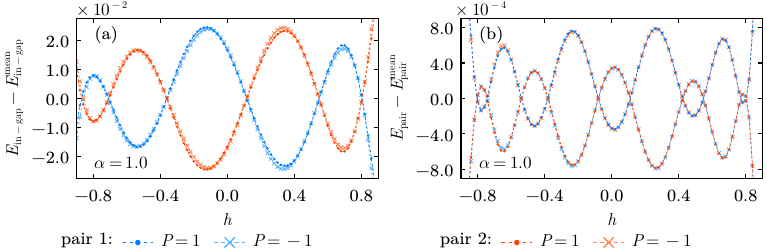}   
    \caption{
        In-gap spectrum of three chains of $N=12$ Majoranas, coupled with a strength $\alpha=1$, in the Y-junction as a function of the magnetic field $h$.
        This spectrum consists of four states that group into two pairs (red and blue data), each pair containing a state with negative (crosses) and positive (dots) parity.
        Four-Majorana interaction strength is fixed at $g=0.2$.
        In both panels, the number of EZM are smaller, and the amplitude of the oscillations bigger, than the Y-junction with three arms of $N=14$ Majoranas each (see Fig.\ref{fig: Y-junction EZM} for reference).
        (a) Spectrum centered around its average.
        Pairs show a vanishing energy gap between them when varying $h$.
        For each pair the two parity sectors are barely distinguishable.
        (b) Same data as in (a) but centered around the average of each parity pair. 
        Opposite parity sectors of both pairs, e.g. red dots and blue crosses, almost perfectly align with each other.
    }
    \label{fig: Y-junction EZM appendix vary h}
  \end{center}
\end{figure}

Characteristic of chains where the interaction between the edge states can be continuously tuned and depends on the distance between the edges is that the amplitude and frequency of the corresponding oscillations depends on the system size \cite{chepigaExactZeroModes2017a,toskovicAtomicSpinchainRealization2016}.
In the Y-junction, these finite size effects also appear.
We show the in-gap spectrum for three arms of $N=12$ Majoranas each in Fig.\ref{fig: Y-junction EZM appendix vary h}(a) and (b). 
The spectrum is qualitatively the same as the one for $N=14$ Majoranas per arm (see Fig.\ref{fig: Y-junction EZM}); the four in-gap states group up into two parity pairs; the energy of these pairs oscillate and the energy gap between them closes; the parity gap within each pair also vanishes frequently.
The amplitude of the oscillations for $N=12$ Majoranas per arm is larger while its frequency is smaller, showing the same finite size effect as for the regular interacting Kitaev chain \cite{chepigaTopologicalQuantumCritical2023a}.

\newpage

\subsection{In-gap spectrum as a function of the four-Majorana interaction strength}\label{Ap: Y-junction vary g}

\begin{figure}[!h]
  \begin{center}
    \includegraphics{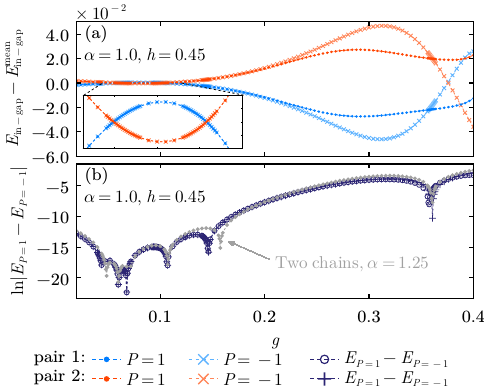}
    \caption{
        Energies of the four in-gap states as a function of the four-Majorana interaction strength $g$ for three chains of $N=14$ Majorana fermions coupled in the Y-junction geometry.
        We fix the magnetic field $h=0.45$ and coupling strength $\alpha=1$.
        (a) Data centered around the average of the in-gap spectrum.
        In-gap states pair up and form two pairs (blue and red curves).
        Dots and crosses denote the positive and negative parity sectors respectively.
        (b) Logarithm of the absolute parity gap $| E_{P=1} - E_{P=-1} |$; dips indicate exact zero modes.
        Data is shown for both parity pairs (dark blue open circles and crosses respectively).
        Grey diamonds correspond to two chains of $N=14$ Majorana fermions each, coupled with $\alpha=1.25$.
    }
    \label{fig: Y-junction EZM vary g}
  \end{center}
\end{figure}

In addition to this, we show the effect varying the four-Majorana interaction strength $g$ on Y-junction in Fig.\ref{fig: Y-junction EZM vary g}(a).
We fix the magnetic field $h=0.45$ and coupling strength $\alpha=1$; the results are centered around their average.
Similar to a single interacting Kitaev chain \cite{chepigaTopologicalQuantumCritical2023a}, we see clear crossings of energy levels upon tuning $g$.
On top of that, the value of $g$ strongly affects the formation of the two parity pairs.
For small values the two parity sectors of each pair show visually inseparable differences. 
On the other hand, this similarity is destroyed for $g \gtrsim 0.2$, resulting in a sequence of pair-wise energy crossings.

We show logarithm of the absolute value of parity gap $E_{P=1} - E_{P=-1}$ for both parity pairs in Fig.\ref{fig: Y-junction EZM vary g}(b).
Dips indicate a vanishing of this gap.
The parity gaps associated to both pairs align closely up to the boundary of the incommensurate region at large $g$.
Rather interestingly, the similarity with the two chains coupled with $\alpha = 1.25$ holds through most of the incommensurate interval. Although the matching coupling $\alpha$ seems non-universal, our results imply that crossings within each parity gap are controlled by outer edge states and can effectively be described by the simplest two-chain junction.

\newpage

\subsection{Spatial profile for all in-gap states}\label{Ap: Y-junction LDOS in-gap states}

\begin{figure}[!h]
  \begin{center}
    \includegraphics{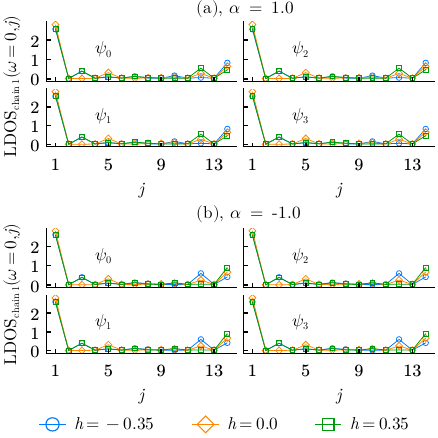}
    \caption{
        Spatial profile of three chains of $N=14$ Majoranas coupled in the Y-junction for the four in-gap states ($\psi_{0,1,2,3}$).
        We show the local density of states for three magnetic fields $h$ (blue circles, orange diamonds, green squares).
        Data is presented for a single arm of the junction only.
        (a) Local density of states for positive coupling $\alpha=1$.
        Note that the profiles are visually identical.
        (b) Same as in (a) but for negative coupling $\alpha=-1$.
    }
    \label{fig: Y-junction LDOS appendix}
  \end{center}
\end{figure}

Similar to the two chains with an impurity bond, the spatial profile of all four in-gap states in the Y-junction shows remarkable similarities. 
In Fig.\ref{fig: Y-junction LDOS appendix}(a) and (b) we show the LDOS of a single arm with $N=14$ Majoranas of the Y-junction for both positive and negative coupling. 
The profiles do not show any visual differences, suggesting that also in more complex structures, such as the Y-junction, the only difference between the states is the interaction between the Majorana degrees of freedom.

\newpage

\section{Additional results X-junction}\label{Ap: D}

\subsection{In-gap spectrum for $N=8$ Majoranas per arm}\label{Ap: X-junction fewer sites}

\begin{figure}[!h]
  \begin{center}
    \includegraphics{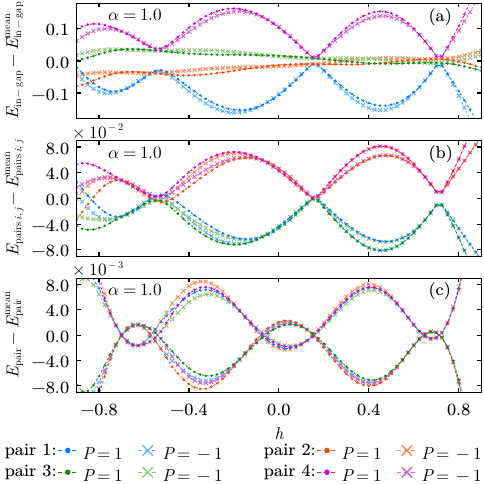}   
    \caption{
        In-gap spectrum for the X-junction with $N=8$ Majoranas per arm as a function of the magnetic field $h$.
        Chains are coupled with a strength $\alpha=1$.
        States group into four pairs (red, blue, green and pink curves); each pair consisting of a state with negative and positive parity, crosses and dots respectively.
        We fix the four-Majorana interaction strength $g=0.2$.
        Amplitude of the oscillations is larger while its frequency is smaller compared to the case of $N=10$ Majoranas per arm (see Fig.\ref{fig: Y-junction EZM}).
        (a) Spectrum centered around its average.
        Red and green data cross once.
        (b) Spectrum of pair one and two centered around its average, and the same for pairs three and four. 
        Only blue and red cross curves cross, green and purple do not. 
        (c) Spectrum of each pair centered around its respective average.
        Parity gap closes in all pairs.
    }
    \label{fig: X-junction EZM appendix}
  \end{center}
\end{figure}

The X-junction shows finite size effects that are expected for systems with edge states where the interaction between the edges can be continuously tuned and is distance dependent \cite{chepigaExactZeroModes2017a,toskovicAtomicSpinchainRealization2016}.
In Fig.\ref{fig: X-junction EZM appendix} we show the spectrum of the eight in-gap states for the X-junction with $N=8$ Majoranas per arm as a function of the magnetic field $h$. 
Qualitatively, the spectrum behaves similar to the one depicted for $N=10$ Majoranas per arm in Fig.\ref{fig: X-junction EZM}.
That is, the states group into four parity pairs; pairs one and two show regular crossings, pair two and three cross only once, and pairs three and four do not cross; the parity gap within each pair closes at approximately the same time.

\newpage

\subsection{Dependence on the four-Majorana interaction strength}\label{Ap: X-junction vary g}

\begin{figure}[!b]
  \begin{center}
    \includegraphics{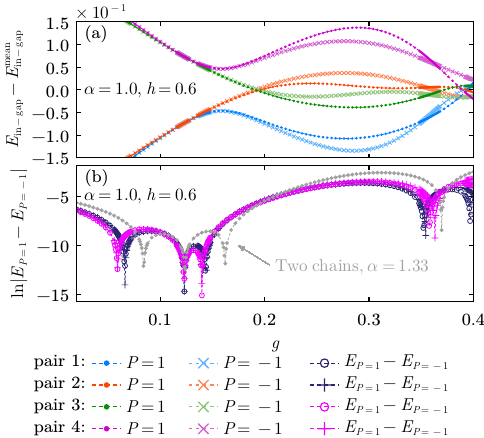}   
    \caption{
        Energy spectrum of the eight in-gap states of four chains coupled in the X-junction geometry as a function of the four-Majorana interaction strength $g$. 
        Data is shown for $N=10$ Majorana fermions per arm, coupled by a strength $\alpha=1.0$, and a fixed magnetic field $h=0.6$.
        States partner up with a single other state with opposite parity $P$ into parity pairs (blue, red, green and pink data).Crosses denote parity $-1$ and dots $+1$.
        (a) Spectrum centered around its average.
        Formation of parity pairs is destroyed for large $g$.
        Third and fourth pair diverge away from the first two for small $g$.
        (b) Logarithm of the absolute value of the parity gap.
        Dips indicate a vanishing of this gap.
        Parity pairs one and two (three and four) are dark purple (neon pink) and odd (even) pairs open circles (crosses).
        For small $g$, the parity gap vanishes simultaneously for the first and second parity pair and, equivalently, for the third and fourth pair, but at a difference location. 
        Parity gap vanishes at difference locations for each pair for $g$ large.
        Grey line with diamonds shows the parity gap for two chains of $N=10$ Majorana fermions coupled by a strength $\alpha=1.33$.
        The two systems show similar behavior.  
    }
    \label{fig: X-junction EZM vary g}
  \end{center}
\end{figure}

We show the dependence of the in-gap spectrum on the four-Majorana interaction strength $g$ in Fig.\ref{fig: X-junction EZM vary g}(a) for $N=10$ Majorana fermions per arm and a fixed magnetic field $h=0.6$.
Similar to the Y-junction, the behavior of the in-gap spectrum strongly depends on the value of $g$.
For small values parity pairs are nicely formed, and there is a clear divergence of the third and fourth pair from the first two as we approach the boundary of the incommensurate region.
In stark contrast to this is the in-gap spectrum at $g \gtrsim 0.18 $.
Similarities between the negative and positive parity sectors in each pair is destroyed and there are a lot of sporadic crossings between the energy levels.
Note, the first and second parity pairs, and the third and fourth, behave similar to those of the Y-junction.

In Fig.\ref{fig: X-junction EZM vary g}(b) we show the logarithm of absolute parity gap of each parity pair. 
A vanishing parity gap occurs at multiple instances of $g$ for each pair. 
Remarkably, only in the region $0.11 \lesssim g \ \lesssim 0.15$ does this vanishing occur at approximately the same point.
For small $g$ the grouping of the first and second, and the third anf fourth parity pairs can be observed: vanishing of the parity gap of the first two pairs occurs simultaneously while it also occurs simultaneously for the third and fourth pair but at a different value of $g$.
On the other boundary vanishing of the gap is much more sporadic and occurs at different locations for each pair. 
We once again observe that the difference within the pairs of eigenvalues in the X-junction is similar to two chains of $N=10$ Majoranas each coupled by a strength $\alpha = 1.33$.

\newpage
\subsection{Spatial profile for all in-gap states}\label{Ap: X-junction LDOS appendix}

\begin{figure}[!h]
  \begin{center}
    \includegraphics[width=\textwidth]{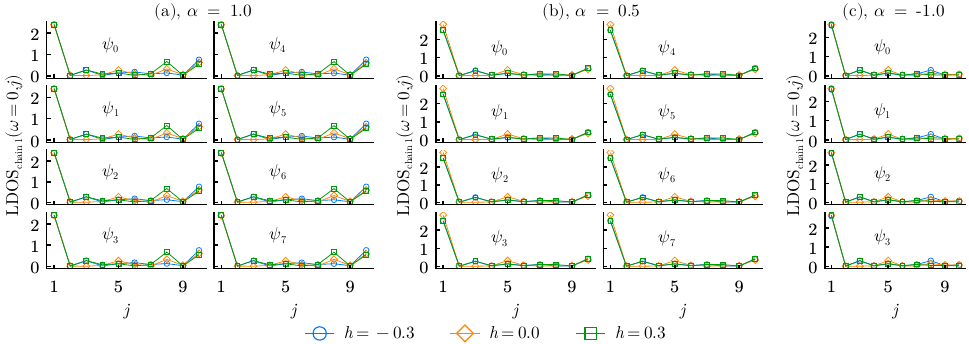}   
    \caption{
        Spatial profile of four chains of $N=10$ Majoranas each coupled at its ends in the X-junction for three different coupling strengths $\alpha$.
        We present the local density of states for a single arm for the in-gap spectrum.
        We show the local density of states for three magnetic fields $h$ (blue circles, orange diamonds, green squares).
        (a) Local density of states for the eight in-gap states ($\psi_{0,1,2,3,4,5,6,7}$) for positive coupling $\alpha=1$.
        Note that the profiles are visually identical.
        (b) Same as in (a) but for coupling $\alpha=0.5$.
        (c) Spatial profiles for the four in-gap states ($\psi_{0,1,2,3}$) for negative coupling $\alpha=-1$.
    }
    \label{fig: X-junction LDOS appendix}
  \end{center}
\end{figure}

We show the LDOS for a single arm of $N=10$ Majoranas of the X-junction in Fig.\ref{fig: X-junction LDOS appendix}(a)-(c) for all in-gap states. 
We present the data for the same couplings as shown in Fig.\ref{fig: X-junction LDOS}.
Again, similar to the other two cases discussed in this study, all in-gap states show identical spatial profiles of the Majoranas, suggesting that these profiles are merely influences by the interaction between the edge states and not the 

\end{appendix}

\newpage

\bibliography{EZM_junction_SciPost}


\end{document}